\documentclass[pra,twoside,amsfonts,amssymb,amsmath,titlepage,floatfix,twocolumn,showpacs,superscriptaddress]{revtex4} 

\usepackage{graphicx}
\usepackage{dcolumn}
\usepackage{bm}

\renewcommand{\phi}{\varphi}
\renewcommand{\>}{\right \rangle}
\newcommand{\<}{\left \langle}

\newcommand{\ket}[1]{\left |#1\>}
\newcommand{\bra}[1]{\<#1\right |}
\newcommand{\ketB}[1]{\big |#1\big >}

\newcommand{\nn}{\nonumber}
\newcommand{\be}{\begin{equation}}
\newcommand{\ee}{\end{equation}}
\newcommand{\bea}{\begin{eqnarray}}
\newcommand{\eea}{\end{eqnarray}}

\newcommand{\Int}{\mathbb{Z}}
\newcommand{\Hr}{\mathfrak{C}_{\mathcal{H}}}
\newcommand{\Real}{\mathbb{R}}
\newcommand{\Comp}{\mathbb{C}}
\newcommand{\Set}{\mathbb{S}}

\def\softt{{\leavevmode\setbox1=\hbox{t}%
\hbox to \wd1{t\kern-0.6ex{\char039}\hss}}}

\begin{document}
\title[Perfect state transfer in networks of arbitrary topology and
interactions]{Perfect state transfer in networks of arbitrary topology and
coupling configuration}

\author{V. Ko\v s\softt\'ak}
 \affiliation{Department of Physics, FJFI \v CVUT, B\v rehov\'a 7, 115 19 Praha 1, Star\'e M\v{e}sto, Czech Republic}

\author{G. M. Nikolopoulos}
\altaffiliation{Present address: Institute of Electronic Structure and 
Laser, FORTH, PO Box 1527, GR-71110 Heraklion, Crete, Greece.}
 \affiliation{Institut f\"ur Angewandte Physik, Technische Universit\"at Darmstadt, 64289 Darmstadt, Germany}

\author{I. Jex}
 \affiliation{Department of Physics, FJFI \v CVUT, B\v rehov\'a 7, 115 19 Praha 1, Star\'e M\v{e}sto, Czech Republic}

\date{\today}

\begin{abstract}
A general formalism of the problem of perfect state transfer is
presented. We show that there are infinitely many Hamiltonians
which may provide solution to this problem. In a first attempt to
give a classification of them we investigate their possible forms
and the related dynamics during the transfer. Finally, we show how
the present formalism can be used for the engineering of perfect
quantum wires of various topologies and coupling configurations.
\end{abstract}


\maketitle

\section{Introduction}
The faithful transfer of a quantum state between two distant but
specified components of a quantum computer is one of the
main requirements for practical quantum computation \cite{divi00}.
The two components (e.g., small quantum processors) are typically
parts of a larger quantum network and connected via a
quantum channel (wire). The state transfer is achieved by
converting a stationary information carrier (qubit)
to a movable (``flying'') one at the input of the wire \cite{divi00}.
The flying qubit is then transmitted through the wire towards its other
end where it is converted back into a stationary qubit.

In general, photons are excellent flying qubits as they can be transmitted
coherently over very large distances \cite{photon}. Hence, many discussions
in the field of quantum computing consider that the state to be
transmitted is first imprinted onto a photon which is used
as the flying qubit over an optical fiber. Clearly, the implementation of
such an idea requires a perfect interface between optical systems and the
main hardware of the quantum computer. However, such an interface is not
always an easy task to realize as the quantum hardware may
be based on atoms, ions, molecules or solid state systems, for example.
Hence, for short-distance communication (e.g., between two quantum processors)
it is desirable to develop new systems which are fully compatible with
the quantum hardware and, in addition, are suitable for faithful
state transfer.

From the point of view of quantum control the problem under consideration
can be rephrased as follows.
Given a set of prescribed elements and gates one has to structure a system
realizing faithful state transfer for arbitrary input states.
Elements of quantum information processing
are typically quantum mechanical objects (e.g., ions, quantum dots,
Josephson junctions, etc) arranged to form linear
chains or planar structures with bipartite interactions which can
be manipulated by certain control parameters. Manipulating the strength
of the interaction between the quantum elements one can design different
types of interactions for the whole system and perform certain gate sequences.

Quantum wires based on systems of permanently coupled quantum
objects are of particular interest as they require minimal
external control thus avoiding significant errors due to the
application of multiple operations (gates) and/or measurements on
various sites of the channel. The design of such a kind of {\em
passive} quantum wires and the problem of faithful state transfer
have attracted considerable interest over the last years. It was
immediately realized, however, that faithful state transfer is not
an easy task, even in the absence of dissipation or dephasing, due
to the dispersion of the quantum information along the wire. Thus,
various methods have been proposed to partially circumvent this
problem and to improve the fidelity of the transfer
\cite{bose03,subra04,ol04,giofa05,RFB05,plesem05,ppkf05,sss05}.

Throughout this work we focus on {\em perfect} state transfer over
passive quantum wires. So far, the existence of perfect passive
quantum wires has been demonstrated in the context of coupled
harmonic oscillators \cite{shore,phe04}, arrays of quantum dots
\cite{NPL04} and spin chains \cite{sss05,CDEKL04,KS05,YB05}. The key
idea is the engineering of certain coupling configurations (and thus
interactions) along the chain which are able to suppress possible
dispersion effects and enforce the complete refocus of the
transmitted quantum information at the ends of the chain at well
defined instants of time. Nevertheless, the majority of these
investigations mainly focus on centrosymmetric (also known as mirror
symmetric) linear chains while the Hamiltonian characterizing the
quantum wire involves nearest-neighbour (NN) interaction only. As a
result, they are not applicable to realistic setups where these
assumptions are relaxed (e.g., beyond NN couplings) \cite{ppkf05},
and the design of new perfect quantum wires  is
necessary \cite{Kay06}.
In the worst case scenario where such a design is not feasible,
numerical optimisation techniques may be invoked for improvement of the
fidelity of the transfer. An alternative solution to this problem relies
on the use of the so-called dual-rail encoding involving two nearly
identical quantum channels \cite{BB05}. However, in this case the
state transfer does not occur at well defined time instants and thus
the arrival of the state can be revealed by means of measurements
only.

The purpose of this paper is manifold. First we aim to solve the
problem of perfect state transfer (PST) in a much broader context.
More precisely, we want to investigate the possible forms that the
system's Hamiltonian may take to be suitable for PST. In contrast to
previous work in the field, here we do not set any {\em a priory}
restrictions to the topology of the system and the configuration of
couplings between different sites of the channel. Hence,
irrespective of topology and coupling configuration, we show that
there are infinitely many Hamiltonians which are suitable for PST.
For the sake of illustration, we demonstrate that certain
Hamiltonians previously discussed in the literature can be obtained
in the framework of our unified theory by setting certain
restrictions and using the right parameterisation. Second, we show
that our approach provides new ways for quantum wire engineering for
systems of arbitrary topology and interactions beyond nearest
neighbours, thus generalising existing work in this context
\cite{KS05,YB05,Kay06}. However, in contrast to
\cite{KS05,YB05,Kay06} our approach to the problem of
PST does not rely on the concept of inverse eigenvalue problems,
but rather on the derivation of Hamiltonians which lead to a particular
transform namely, a permutation operation.
Finally, the present work can also be viewed as an attempt to
classify the Hamiltonian problems leading to exact revivals in
discrete time evolution and the types of dynamics that appear during
this evolution.

Our theoretical approach was also used for the derivation of Hamiltonians
for the Fourier transform implemented using linear optical
elements \cite{TJS}. The general form of the Hamiltonian was
given and the results are closely linked to the present ones as the
Fourier transform is a cyclic operation (a property shown to be
crucial for our considerations). The square of the Fourier
transform gives a permutation operation, hence is related to the
problem under investigation in the following sections.

The paper is organised as follows. In Sec. \ref{Sec2} we introduce
our mathematical model and derive a class of Hamiltonians suitable for
PST. In Sec. \ref{Sec3}, considering small networks we demonstrate
how our theoretical approach can be used for quantum wire engineering.
We show that certain Hamiltonians previously discussed in the literature
are  members of the larger unifying class derived in Sec. \ref{Sec2}.
Moreover, we derive new PST Hamiltonians in the framework of
NN-type, coulomb and dipole-dipole interactions.  In Sec. \ref{Sec4},
we analyse and discuss the dynamics induced by various types of PST
Hamiltonians while we conclude with a summary of our main results in
Sec. \ref{Sec5}.

\section{Hamiltonians for perfect state transfer}
\label{Sec2}

The model we consider is a network consisting of $n$ sites
labelled by $\{1, 2, \ldots, n\}\equiv\Set_n$. At time $t=0$, the
qubit (e.g, spin) in the $1$-st (input) site of the network is
prepared in the state $\ket{\psi_{\rm in}}$. We wish to transfer
to the $n$-th (output) site of the network with unit efficiency
after a well defined period of time, let us say $t_{\rm r}\equiv
\tau/J$, where $J^{-1}$ is our time units $(\hbar=1)$.

We are interested in perfect passive quantum wires i.e., networks
involving permanently coupled sites without any additional
external control. Depending on the physical realization of the
network, $J$ can be a characteristic energy in our system, a
coupling constant or tunnelling (hopping) rate, etc. Nevertheless,
to present a generic theoretical approach to the problem of PST,
the following discussion will be based on the dimensionless
quantity $\tau$. It is also worth noting, that $\tau$ need not
always be a continuous variable as in some physical realizations
(e.g., passive linear optical networks), the excitation (i.e.,
photon) evolves under successive applications of identical
(practically instantaneous) unitary operations induced by the
system's Hamiltonian \cite{single}. In this case $\tau$ is simply
the number of unitary operations we have applied. The subsequent
discussion applies to all these physical realizations no matter
whether they involve discrete or continuous evolution. However,
for the sake of brevity we mainly refer to $\tau$ as number of
applications (iterations) of identical unitary operations.

Our purpose is to explore the possible forms that the system's
Hamiltonian ${\cal H}$ may take, to be suitable for PST. In other
words, we wish to analyse the whole class of Hamiltonians $\Hr$,
which lead to PST from the first to the last site after exactly
$\tau$ applications. Following previous work in the field
\cite{sss05,shore,phe04,NPL04,CDEKL04,KS05,YB05,Kay06,BB05}, we
will assume that the network does not disturb the transmitted
state and the main source of possible dissipation and decoherence
is the fact that the wave packet (quantum information) spreads
along the network. Hence, we are interested in Hamiltonians which
preserve the total number of excitations in the system. 
For instance, in the framework of spin
chains and arrays of quantum dots this means that the Hamiltonian
commutes with the total spin operator
\cite{NPL04,CDEKL04,KS05,YB05,Kay06,BB05}.

One way to guarantee all of these requirements is to concentrate our
subsequent investigation on
Hamiltonians for which the associated unitary evolution ${\cal U}$,
leads to a permutation matrix after $\tau$ applications i.e.,
\bea
{\cal U}(\tau)\equiv e^{{\rm i}{\cal H}\tau} = {\cal P}, \label{eq:1eq}
\eea
where
\bea
{\cal P}=\left(
\begin{array}{cccc}0& & &\\\vdots& &\tilde{{\cal P}} &\\0& & & \\1&0&\cdots&0
\end{array}
\right) , \label{eq:generalP} \eea is a permutation in the single excitation sector i.e. also the
submatrix $\tilde{\cal P}$ in the computational basis is a
permutation. The permutation matrix (\ref{eq:generalP}) guarantees
that the net effect of the evolution (\ref{eq:1eq}) is the transfer of the
excitation from the first position to the last one i.e.,
$\vert 1\rangle_1
\vert 0\rangle_2 ...\vert 0\rangle_n\rightarrow \vert 0\rangle_1
\vert 0\rangle_2 ...\vert 1\rangle_n$. The condition
(\ref{eq:1eq}) might seem restrictive at this point and this is
in general indeed the case as one may derive Hamiltonians which
satisfy the aforementioned requirements and do not lead to
permutations, but rather to other unitary operations. Nevertheless,
as we show later on, many known Hamiltonians suitable for PST are
basically associated with permutations and can thus be obtained
within the present unifying theoretical framework. Besides, we will
demonstrate that we have infinitely many, yet unexplored, choices
for quantum wire engineering which are not covered by previous work
in the field. We have to point out, however, that condition
(\ref{eq:1eq}) is indeed a severe restriction if one is
interested only in perfect transfer of a particular set of possible
states or properties of states such as probabilities, entanglement,
etc.

Note now that perfect transfer of arbitrary single-qubit states
is a sufficient condition for perfect transfer of arbitrary 
multi-qubit states. For instance, in this case the same perfect 
quantum wire can be used multiple 
times for the transfer of each qubit separately. Furthermore,
as long as the number of excitations and the transmitted qubit
state are preserved by the Hamiltonian, the total Hilbert space in
the problem can be decomposed into subspaces and  we can focus on
the one-excitation subspace. The corresponding basis is denoted by
$\left \{\ket{\alpha}~|~\alpha\in\Set_n\right \}$ and indicates
the presence of the excitation at the site $\alpha$. Hence, the
problem of the PST is essentially reduced to the perfect transfer
of a single excitation initially located at the $1$-st site i.e.,
$\ket{\psi_{\rm in}}=\ket{1}$.

In general, we can define $(n-1)!$ different permutations of the form
(\ref{eq:generalP}), we have $n-1$ free positions to fill, and for each
one of these our purpose now is to construct the class
of Hamiltonians satisfying Eq.~(\ref{eq:1eq}) and thus are suitable
for PST. We can distinguish between cases where ${\cal P}$ consists
of one or more cycles. In the following, we are going to investigate
the two cases separately but we will see that the latter reduces to
the former one within each cycle.

\subsection{One-cycle permutations}
\label{Sec2.1}
For $n$ sites we have $(n-2)!$ possible one-cycle permutations.
In particular, starting from the $n\times n$ permutation
 \bea
{\cal P}=\left(
\begin{array}{ccccc}0&1&\cdots&0\\
\vdots&\ddots&\ddots&\vdots\\0&\cdots &0&1\\1&0&0&0
\end{array}
\right), \label{eq:1cycle} \eea the other possible one-cycle
permutations can be obtained by simply relabelling the $(n-2)$
intermediate sites $\{2, \ldots,n-1\}$. In the following we will
derive our results for this particular matrix.

To determine the class of Hamiltonians $\Hr$ satisfying
Eq.~(\ref{eq:1eq}) for the permutation (\ref{eq:1cycle}), we need to
know the eigenvalues and eigenvectors of the operator ${\cal P}$.
The spectrum $\sigma$ consists of $n$ different eigenvalues of the
characteristic equation $\lambda^n=1$ i.e., \be
\sigma=\{\lambda_j~|~\lambda_i\neq\lambda_j~\textrm{for}~i,j\in\Int_n\},
\ee where \bea \lambda_j=\exp\left ({\rm i}2\pi\frac{j}{n}\right
)\quad\textrm{for} \quad j\in\Int_n, \label{eq:1cycleSpectrum} \eea
and $\Int_n\equiv\{0,1,\dots,n-1\}$. The corresponding normalised
eigenvectors can be expanded in the computational basis
$\{\ket{\alpha}~|~\alpha\in\Set_n\}$ as follows \bea
\ket{y_{\lambda_j}}=\frac1{\sqrt{n}}
\sum_{\alpha\in\Set_n}\lambda_j^{\alpha-1}
\ket{\alpha}=\frac1{\sqrt{n}}\left(1,\lambda_j^1,\ldots,\lambda_j^{n-1}\right),
\label{eq:1cycleEVector} \eea and thus the spectral representation
of ${\cal P}$ is \bea {\cal P}
=\sum_{\lambda_j\in\sigma}\lambda_j\ket{y_{\lambda_j}}\bra{y_{\lambda_j}}.
\label{eq:SpecP}
\eea

Using Eqs.~(\ref{eq:1cycleSpectrum})-(\ref{eq:SpecP}), it is
straightforward to construct a first Hamiltonian which satisfies
Eq.~(\ref{eq:1eq}) and (\ref{eq:1cycle}) as follows \bea
\label{eq:ham}
H=\frac{1}{\tau}\sum_{\lambda_j\in\sigma}\arg(\lambda_j)
\ket{y_{\lambda_j}}\bra{y_{\lambda_j}}
\eea where $\arg(\lambda_j)$ is the phase of the $j$-th eigenvalue
$\lambda_j$. However, this is not the only Hamiltonian which leads
to the permutation ${\cal P}$ after $\tau$ applications.
Additionally, we may shift each eigenenergy of the Hamiltonian
(\ref{eq:ham}) by an arbitrary integer multiple of $2\pi$
obtaining
\bea {\cal H}_\mathbf{l}&=&\frac1\tau
H+\sum_{\lambda_j\in\sigma} \frac{2\pi
l_{\lambda_j}}{\tau}\ket{y_{\lambda_j}}\bra{y_{\lambda_j}}\nonumber\\
&=&\frac{1}{\tau}\sum_{\lambda_j\in\sigma}\left [\arg(\lambda_j)+2\pi
l_{\lambda_j}\right ]
\ket{y_{\lambda_j}}\bra{y_{\lambda_j}},\label{eq:hame}
\eea
with $\mathbf{l}\in\Int^n\equiv\{(l_{\lambda_0},l_{\lambda_1},\ldots,
l_{\lambda_{n-1}})~|~l_{\lambda_j}\in\Int,j\in\Int_n\}$. Therefore
the class of Hamiltonians satisfying Eq.~(\ref{eq:1eq}) for the
particular one-cycle permutation (\ref{eq:1cycle}) is \bea \Hr =
\{{\cal H}_\mathbf{l}~|~\mathbf{l}\in\Int^n\}, \label{eq:class} \eea
with ${\cal H}_\mathbf{l}$ given by Eq.~(\ref{eq:hame}). Since the
members of $\Hr$ are parameterised by the integer vector
$\mathbf{l}\in\Int^n$, the class  $\Hr$ consists of infinitely many
Hamiltonians suitable for PST.

Note now that all the members of $\Hr$ are linear superpositions of
the projectors onto the eigensubspaces of the permutation
${\cal P}$. More precisely, Eq.~(\ref{eq:hame}) can be also written as \bea
{\cal H}_\mathbf{E}=\sum_{\lambda_j\in\sigma}
\varepsilon_{\lambda_j}\ket{y_{\lambda_j}}\bra{y_{\lambda_j}}
\equiv\sum_{\lambda_j\in\sigma} \varepsilon_{\lambda_j}\Pi_{\lambda_j},
\label{gen_Ham_diag} \eea where the {\em spectrum}
(eigenenergy-vector) $\mathbf{E}\in\Real^n\equiv
\{(\varepsilon_{\lambda_0},\ldots,
\varepsilon_{\lambda_{n-1}})~|~\varepsilon_{\lambda_j} \in\Real,
j\in\Int_n\}$.

Using Eq.~(\ref{eq:1cycleEVector}), we can express
Eq.~(\ref{gen_Ham_diag}) in the computational basis as follows \bea
{\cal H}_\mathbf{E}=\sum_{\alpha,\beta\in\Set_n}
\sum_{\lambda_j\in\sigma}
\varepsilon_{\lambda_j}\lambda_j^{\alpha-\beta}
\ket{\alpha}\bra{\beta}, \eea which can be also written in the usual
form \be {\cal H}_\mathbf{E}={\cal H}_\mathbf{E}^{(0)}+{\cal
V}_{\mathbf{E}} \label{gen_Ham_dec} \ee with the diagonal part \be
\label{gen_Ham_unp} {\cal
H}_\mathbf{E}^{(0)}=\sum_{\alpha\in\Set_n}E_{\alpha}
\ket{\alpha}\bra{\alpha}\ee
 and the interaction\be
{\cal V}_{\mathbf{E}}=\sum_{\alpha\neq\beta\in\Set_n}
G(\alpha,\beta) \ket{\alpha}\bra{\beta}. \label{gen_Ham_int} \ee
Thereby, the energies and the couplings are given by \bea
\label{gen_Ham_en}
E_{\alpha} &=& \sum_{\lambda_j\in\sigma} \varepsilon_{\lambda_j},\\
G(\alpha,\beta) &=& \sum_{\lambda_j\in\sigma}
\varepsilon_{\lambda_j}\lambda_j^{\alpha-\beta}.
\label{gen_Ham_coup} \eea From Eq.~(\ref{gen_Ham_en}), we see that
Hamiltonians which are associated with permutation
(\ref{eq:1cycle}) and lead to PST may only correspond to
networks involving the same energy level for all the sites.

The conditions under which  Hamiltonians with NN interaction may lead to
PST have been extensively discussed in the literature during the last
years \cite{sss05,phe04,NPL04,CDEKL04,KS05,YB05}.
Such NN-interaction Hamiltonians are typically tridiagonal in the
computational basis i.e., we have $G(\alpha,\beta)=0$, for all
$\beta\notin\{\alpha,\alpha\pm 1\}$. However, according to the
following theorem the class of Hamiltonians $\Hr$ we have
derived here does not include NN-interaction Hamiltonians.
\\ \\
{\bf Theorem.} {\em For networks of arbitrary dimension $(n>2)$,
there exists no nearest-neighbour-interaction Hamiltonian satisfying
condition (\ref{eq:1eq}) in the framework of permutation (\ref{eq:1cycle})}.
\\ \\
{\bf Proof.} According to Eq.~(\ref{gen_Ham_diag}) all the
Hamiltonians which satisfy Eq.~(\ref{eq:1eq}) with ${\cal P }$ given
by Eq.~(\ref{eq:1cycle}) are linear superpositions of the projectors
$\Pi_j$. Moreover, they can be decomposed as described in
Eqs.~(\ref{gen_Ham_dec})-(\ref{gen_Ham_coup}). For a given dimension
$n>2$, let us now assume that there exists a set of eigenenergies
$\mathbf{x}\in\Real^n$ such that the corresponding Hamiltonian
${\cal H}_\mathbf{x}$ involves NN interaction. We are
going to show that such a Hamiltonian does not exist (proof by
contradiction).

We start with the general decomposition of the interaction term
${\cal V}_\mathbf{x}$ as \bea {\cal V}_\mathbf{x}={\cal
V}_\mathbf{x}^{({\rm N})}+ {\cal V}_\mathbf{x}^{(\not{\rm N})} ,
\eea where ${\cal V}_\mathbf{x}^{({\rm N})}$ involves NN
interaction i.e., $\beta=\alpha\pm 1$, while ${\cal
V}_\mathbf{x}^{(\not{\rm N})}$ involves terms with
$\beta\notin\{\alpha,\alpha\pm 1\}$. Our assumption about the
existence of an NN-type Hamiltonian implies that the
Hamiltonian is tridiagonal in the computational basis, i.e., \be
{\cal V}_\mathbf{x}^{({\rm N})}\neq 0\quad \textrm{and}\quad{\cal
V}_\mathbf{x}^{(\not{\rm N})} = 0. \label{NN-cond} \ee From the
second condition and Eq.~(\ref{gen_Ham_int}) we have that
$G(\alpha,\beta)=0$, for all $\beta\notin\{\alpha,\alpha\pm 1\}$.
However, since $G(\alpha,\beta)$ depend only on $\alpha-\beta$ we
may obtain $n-2$ linear equations for variables $x_j$ (see
Eq.~\ref{gen_Ham_coup}), namely $\Lambda \mathbf{x}=\mathbf{0}$
where \bea \Lambda=\left(\begin{array}{cccc}
\lambda_0^2&\lambda_1^2&\cdots&\lambda_{n-1}^2\\
\lambda_0^3&\lambda_1^3&\cdots&\lambda_{n-1}^3\\
\vdots& & &\vdots\\
\lambda_0^{n-1}&\lambda_1^{n-1}&\cdots&\lambda_{n-1}^{n-1}
\end{array}\right).
\label{eq:NN1cycle}
 \eea
Note now that, in view of Eq.~(\ref{eq:1cycleSpectrum}), the matrix
(\ref{eq:NN1cycle}) is (up to first two missing rows and
normalisation) the usual Fourier transformation. Hence,
$\textrm{rank}(\Lambda)=n-2$, while two linear independent solutions
can be chosen as $\mathbf{\tilde x}_1=(1,\ldots,1)$ and
$\mathbf{\tilde x}_2=(\overline{\lambda}_0,\ldots,
\overline{\lambda}_{n-1})$. Asking for a linear combination of
$\mathbf{\tilde x}_1$ and $\mathbf{\tilde x}_2$ to be real (because
this plays the role of eigenenergy vector) we find that the only
acceptable solution is multiple of $(1,\ldots,1)$. However, one can
see immediately from Eqs.~(\ref{gen_Ham_unp}) and
(\ref{gen_Ham_int}), that such a spectrum can be associated only
with the system involving no interaction between sites i.e., \be
{\cal V}_\mathbf{x}^{({\rm N})}= 0\quad \textrm{and}\quad{\cal
V}_\mathbf{x}^{(\not{\rm N})} = 0. \ee which contradicts our initial
assumption (\ref{NN-cond})
about NN interaction. $\Box$\\

Closing the section we would like to emphasise that the presented
method (including the proof) applies also to all other one cycle
permutations. The explicit results will differ from the presented
ones only by the corresponding permutation of the labels.

\subsection{Many-cycle permutations}
\label{sec:manyCycles}

In the case of permutations consisting of more than one cycle, the
situation is slightly more complicated but the problem can be
treated separately within each cycle along the lines of the
previous section. More precisely, consider a cycle of length
$d<n$. Such a cycle does not involve all the network sites but
rather a subset of them $\Set_d\subset\Set_n$. Hence, the related
eigenvalues are given by \bea \lambda_j=\exp\left ({\rm
i}2\pi\frac{j}{d}\right )\quad\textrm{for} \quad j\in\Int_d,
\label{eq:McycleSpectrum}\eea while for the projections of the
corresponding eigenvectors onto the computational basis
$\{\ket{\alpha}~|~\alpha\in\Set_n\}$ we have
\bea
\Big\langle\alpha\Big|v_{\lambda_j}^{(k)}\Big\rangle=
\left\{\begin{array}{ll}
\lambda_j^{\alpha-1}/\sqrt{d} & \textrm{for $\alpha\in\Set_d$}\\
0 & \textrm{otherwise}
\end{array}\right . .
\label{eq:McycleEVector} \eea As is apparent from
Eq.~(\ref{eq:McycleSpectrum}) the spectrum $\sigma$ of a many-cycle
permutation is always degenerate since at least one of the
eigenvalues (i.e., the eigenvalue for $j=0$) appears as many times
as the total number of cycles. Hence, the same degenerate
eigenvalue $\lambda_j$ corresponds to $\delta_{\lambda_j}$ distinct
eigenvectors from different cycles. To this end, in
Eq.~(\ref{eq:McycleEVector}) the eigenvectors are characterised by
the additional superscript $k\in\{1,\ldots,\delta_{\lambda_j}\}$,
where $\delta_{\lambda_j}$ is the degeneracy of the eigenvalue
$\lambda_j$. For a given degenerate eigenvalue $\lambda_j$, let us
also denote by ${\cal E}_{\lambda_j}$ the subspace spanned by the
$\delta_{\lambda_j}$ distinct eigenvectors
$\{|v_{\lambda_j}^{(k)}\rangle\}$.

In view of the degeneracy in the spectrum of a many-cycle
permutation we have more freedom in the construction of Hamiltonians
satisfying condition (\ref{eq:1eq}). Indeed, as described in
Sec. \ref{Sec2.1},
for such a construction we first of all need a basis
of orthonormal eigenvectors. However, within each eigensubspace
${\cal E}_{\lambda_j}$ we can choose such an eigenbasis
$\{|y_{\lambda_j}^{(k)}\rangle\}$ in many different ways, by
constructing linear superpositions of
$\{|v_{\lambda_j}^{(k)}\rangle\}$. Moreover, for each eigenvector
$|y_{\lambda_j}^{(k)}\rangle$ we have the additional freedom to
shift the phase of its eigenvalue by a multiple of $2\pi$, as we did
in Sec. \ref{Sec2.1}. So, the class of PST Hamiltonians (\ref{eq:class})
has infinitely many members of the form
\bea {\cal
H}_{\mathbf{l}}&=&\frac1\tau\sum_{\lambda_j}
\sum_{k=1}^{\delta_{\lambda_j}}\left [ \arg\left (\lambda_j\right )+ 2\pi
l_{\lambda_j}^{(k)}\right ]
\ket{y_{\lambda_j}^{(k)}}\bra{y_{\lambda_j}^{(k)}}\nn\\&=&\frac1\tau\sum_{\lambda_j}
\sum_{k=1}^{\delta_{\lambda_j}}\varepsilon_{\lambda_j}^{(k)}
\ket{y_{\lambda_j}^{(k)}}\bra{y_{\lambda_j}^{(k)}},
\label{deg_Ham}
\eea where
$\mathbf{l}\in\Int^n\equiv\{(l_{\lambda_0}^{(1)},\ldots,l_{\lambda_0}^{(\delta_{\lambda_0})};l_{\lambda_1}^{(1)},\ldots,l_{\lambda_1}^{(\delta_{\lambda_1})};\ldots)~|~l_{\lambda_j}^{(k)}\in\Int\}$.
This is a straightforward generalisation of Eq.~(\ref{eq:hame}) to
the case of degenerate eigenvalues.

\subsection{Quantum wire engineering} So far we have seen that given
a permutation matrix of a particular form one may easily derive
the whole class of PST Hamiltonians which lead to the
permutation under consideration. Hence, our method provides a way of
quantum wire engineering in the sense that it enables us to construct
possible PST Hamiltonians (i.e., to define energies and coupling strengths)
which are implementable in a particular setup. The starting point is always
the choice of the permutation which need not be always random,
as various factors (such as topology of our network and type of interactions)
may automatically fix our permutation matrix. For instance, in the case
of a centrosymmetric network one has to focus on antidiagonal permutation
matrices only while, according to the theorem of Sec. \ref{Sec2.1},
NN-type Hamiltonians automatically exclude all one-cycle permutations.

Having fixed our permutation, the construction of
a PST Hamiltonian proceeds in two steps. In the first step one has
to define the corresponding class of PST Hamiltonians working along
the lines of this section. The members of this class are parameterised
by a number of free parameters. In the second step one
can estimate all these parameters by applying certain constraints
based on the topology of the network
as well as the form of the physical interactions implementable within
the framework of a particular setup.
This engineering process can be always performed numerically but in
certain cases (especially for relatively small networks) derivation
of analytic solutions might be possible. In the following section
we discuss the quantum wire engineering in detail by explicitly
constructing PST Hamiltonians for small networks.

\section{Examples of PST Hamiltonians}
\label{Sec3}
Let us devote some attention to the application of the proposed
quantum wire engineering method to few concrete examples.
Among the simplest we can think of, is the example of a small
networks consisting of just few sites.

\subsection{Nearest-neighbour interaction}
Following the notation introduced in the previous section,
we start with the investigation of the whole class of NN-type PST
Hamiltonians in the context of a small network consisting of
four sites only. The problem of PST in such a small network
is amenable to analytic solutions thus offering the appropriate
theoretical framework to demonstrate the application of our method.

We can define six $4\times 4$ permutation matrices of the form
(\ref{eq:generalP})
that is, two one-cycle permutations and four permutations
involving more than one cycles. As we showed in the previous section,
NN-type Hamiltonians cannot be obtained in the context of
one-cycle permutations. Hence, for the purposes of this particular example,
it is sufficient to restrict ourselves
to permutation matrices involving more than one cycles only.

Let us consider first the antidiagonal permutation matrix
\bea {\cal P}_4=\left (
\begin{array}{ccccccc}
0 & 0 & 0 & 1\\
0 & 0 & 1 & 0 \\
0 & 1 & 0 & 0  \\
1 & 0 & 0 & 0
\end{array}
\right ).
\label{anti_perm}
\eea This permutation involves two cycles of dimension
$d=2$ namely, ${\cal C}_1=(1,4)$ and ${\cal C}_2=(2,3)$
\footnote{Thereby the notation $(2,3)$ means that starting from
the original site ordering $\{1,2,3,4\}$, the second site is
replaced by the third and the third site by the second.}. Hence,
according to Eq. (\ref{eq:McycleSpectrum}) we have two
eigenvalues $\lambda_0=e^{{\rm i}0}=+1$ and $\lambda_1=e^{{\rm
i}\pi}=-1$. Both eigenvalues $\pm 1$ are doubly degenerate (i.e.,
$\delta_\pm=2$) and the corresponding eigenvectors are given by
$\ketB{v_{\pm}^{(1)}}=(\ket{1}\pm\ket{4})/\sqrt{2}$ and
$\ketB{v_{\pm}^{(2)}}=(\ket{2}\pm\ket{3})/\sqrt{2}$, for cycles
${\cal C}_1$ and ${\cal C}_2$, respectively. Accordingly, the
corresponding subspaces are
\[{\cal E}_\pm=\left\{\ket{v_\pm^{(1)}},\ket{v_\pm^{(2)}}\right\}.\]
Having defined the permutation matrix, the construction of PST
Hamiltonians now proceeds in two steps.

\subsubsection{Step 1 --- Parameterisation.}
For each one of the subspaces we can choose an orthonormal basis in
many different ways. For example, taking into
account the orthogonality condition, we may choose for the two subspaces
\bea
{\cal E}_+: \left \{
\begin{array}{l}
\ket{y_+^{(1)}}=\nu\ket{v_+^{(1)}}+\mu \ket{v_+^{(2)}}\\ \\
\ket{y_+^{(2)}}=\mu^* \ket{v_+^{(1)}}-\nu^* \ket{v_+^{(2)}},
\end{array}\right.
\eea
and
\bea
{\cal E}_-: \left \{ \begin{array}{l}
\ket{y_-^{(1)}}=\xi
\ket{v_-^{(1)}}+\zeta \ket{v_-^{(2)}}\\ \\
\ket{y_-^{(2)}}=\zeta^* \ket{v_-^{(1)}}-\xi^* \ket{v_-^{(2)}}.
\end{array}\right.
\eea
Thereby, $\mu,\nu,\xi,\zeta\in\Comp$ such that
\be
|\mu|^2+|\nu|^2=1\quad\textrm{and}\quad |\xi|^2+|\zeta|^2=1.
\label{norm_cond}
\ee
According to the theory of the previous section, the  corresponding
eigenergies can be chosen as
\bea
{\cal E}_+: \left \{
\begin{array}{l}
\varepsilon_{+}^{(1)}= 0+2\pi l_+^{(1)}\\ \\
\varepsilon_{+}^{(2)}= 0+2\pi l_+^{(2)},\end{array}\right.
\eea
and
\bea
{\cal E}_-: \left \{ \begin{array}{l}
\varepsilon_{-}^{(1)}= \pi+2\pi l_-^{(1)}\\ \\
\varepsilon_{-}^{(2)}= \pi+2\pi l_-^{(2)},
\end{array}\right.
\eea
for $l_j^{(k)}\in\Int$. Thereby, note that the eigenvalues
of subspace ${\cal E}_+$ (${\cal E}_-$) are even (odd)
integer multiples of $\pi$.

Thus, the whole class of PST Hamiltonians (\ref{deg_Ham}) reads
\bea {\cal H}_{\mathbf{l}}^{(4)}&=&\frac{1}{\tau}\bigg [
\varepsilon_+^{(1)}\ket{y_+^{(1)}}\bra{y_+^{(1)}}+
\varepsilon_+^{(2)}\ket{y_+^{(2)}}\bra{y_+^{(2)}}\nonumber\\
&&+\varepsilon_-^{(1)}\ket{y_-^{(1)}}\bra{y_-^{(1)}}+
\varepsilon_-^{(2)}
\ket{y_-^{(2)}}\bra{y_-^{(2)}}\bigg ],
\label{NN4_class}
\eea
with open parameters the spectrum
$\big \{\varepsilon_{+}^{(1)},\varepsilon_{+}^{(2)},\varepsilon_{-}^{(1)},
\varepsilon_{-}^{(2)}\big \}$
[or equivalently the integers $l_\pm^{(1,2)}$]
and two independent complex numbers (due to normalisation),
say $\mu$ and $\xi$. As we discuss in the following
subsection, all these parameters can be
specified by imposing additional constraints on the form of the
resulting Hamiltonian. Moreover, we would like to note that, due to the form of
the eigenvectors, all the members of this class are symmetric along the diagonal
and the antidiagonal.

\subsubsection{Step 2 --- Parameter estimation.}
Assume now that we are interested in the whole class of NN-type
Hamiltonians leading
to the antidiagonal permutation matrix (\ref{anti_perm}) and are
of the form
\bea {\cal H}^{(4)}=\left (
\begin{array}{cccc}
E_1 & g_1 & 0 & 0\\
g_1 & E_2 & g_2 & 0 \\
0 & g_2 & E_2 & g_1  \\
0 & 0 & g_1 & E_1
\end{array}
\right ).
\label{eq:ham4}
\eea
Since the class (\ref{NN4_class}) has been derived within a rather
general framework (limited only by the particular form of the
permutation), our task reduces to the application of specific
constraints on the parameters entering ${\cal H}_{\bf l}^{(4)}$.

We will focus on the case of Hamiltonians with non-degenerate spectrum
i.e., all the eigenenergies $\varepsilon_\pm^{(1,2)}$ are different.
The case of degenerate spectrum can be treated similarly and leads
to NN-type Hamiltonians with vanishing couplings. Hence, such solutions
correspond to a broken network and can never lead to PST from the first
to the last site.\\

Rewriting Eq. (\ref{NN4_class}) in the computational
basis and asking for Hamiltonians of the form (\ref{eq:ham4}),
we have that the matrix elements $\bra{1}{\cal H}_{\bf l}^{(4)}\ket{3}$
and  $\bra{1}{\cal H}_{\bf l}^{(4)}\ket{4}$ must vanish.
Thus we obtain the following set of nontrivial constraints on the free parameters,
\begin{subequations}
\label{NN4all}
\bea
\label{NN4:1}
&&\varepsilon_{+}^{(1)}|\nu|^2+\varepsilon_{+}^{(2)}|\mu|^2-
\varepsilon_{-}^{(1)}|\xi|^2-\varepsilon_{-}^{(2)}|\zeta|^2=0,
\\
\label{NN4:2}
&&(\varepsilon_{+}^{(1)}-\varepsilon_{+}^{(2)})\nu\mu^*
+(\varepsilon_{-}^{(2)}-\varepsilon_{-}^{(1)})\xi\zeta^*=0.
\eea
\end{subequations}
Other equations which can be obtained from the remaining matrix elements
$\bra{i}{\cal H}_{\bf l}^{(4)}\ket{j}$ will be used later on, for the
derivation of sets $\{E_1,E_2\}$ and  $\{g_1,g_2\}$ for which the
Hamiltonian (\ref{eq:ham4}) enables PST.

Using the normalisation conditions (\ref{norm_cond}), Eq. (\ref{NN4:1})
can be also expressed in terms of differences of eigenvalues only i.e.,
$$
(\varepsilon_{+}^{(1)}-\varepsilon_{+}^{(2)})|\nu|^2
-(\varepsilon_{-}^{(1)}-\varepsilon_{-}^{(2)})|\xi|^2+
\varepsilon_{+}^{(2)}-\varepsilon_{-}^{(2)}=0.
$$
This fact indicates that the existence of a solution to the problem
of PST depends mainly on differences of eigenenergies. In other
words, if we have a solution for a particular choice of the
spectrum, say
$\{\varepsilon_1,\varepsilon_2,\varepsilon_3,\varepsilon_4\}$, then
there also exist the solutions for all the spectra of the form
$\{\varepsilon_1+2\pi j,\varepsilon_2+2\pi j,\varepsilon_3+2\pi
j,\varepsilon_4+2\pi j\}$ with $j \in \Int$.

Assume now that $\nu$ and $\xi$ are real and non-negative. Such an
assumption is not restrictive because the global phase of each
eigenvector can be chosen arbitrarily without changing the
Hamiltonian (and thus dynamics). The phases of $\mu$ and $\zeta$
will be denoted by $\phi$ and $\chi$, respectively. Solving Eqs.
(\ref{NN4all}) with the normalisation conditions (\ref{norm_cond}),
we obtain for the absolute values of the parameters $\mu$, $\nu$,
$\xi$, and $\zeta$ \bea
 |\mu|&=&\sqrt{\frac{(\varepsilon_{+}^{(2)}-\varepsilon_{-}^{(2)})(\varepsilon_{-}^{(1)}-\varepsilon_{+}^{(2)})}{(\varepsilon_{+}^{(1)}-\varepsilon_{+}^{(2)})(\varepsilon_{+}^{(1)}-\varepsilon_{-}^{(2)}-\varepsilon_{-}^{(1)}+\varepsilon_{+}^{(2)})}},\nn\\
|\nu|&=&\sqrt{\frac{(\varepsilon_{-}^{(1)}-\varepsilon_{+}^{(1)})(\varepsilon_{-}^{(2)}-\varepsilon_{+}^{(1)})}{(\varepsilon_{+}^{(1)}-\varepsilon_{+}^{(2)})(\varepsilon_{+}^{(1)}-\varepsilon_{-}^{(2)}-\varepsilon_{-}^{(1)}+\varepsilon_{+}^{(2)})}},\nn\\
|\xi|&=&\sqrt{\frac{(\varepsilon_{-}^{(2)}-\varepsilon_{+}^{(1)})(\varepsilon_{-}^{(2)}-\varepsilon_{+}^{(2)})}{(\varepsilon_{-}^{(1)}-\varepsilon_{-}^{(2)})(\varepsilon_{+}^{(1)}-\varepsilon_{-}^{(2)}-\varepsilon_{-}^{(1)}+\varepsilon_{+}^{(2)})}},\nn\\
|\zeta|&=&\sqrt{\frac{(\varepsilon_{-}^{(1)}-\varepsilon_{+}^{(1)})(\varepsilon_{-}^{(1)}-\varepsilon_{+}^{(2)})}{(\varepsilon_{-}^{(2)}-\varepsilon_{-}^{(1)})(\varepsilon_{+}^{(1)}-\varepsilon_{-}^{(2)}-\varepsilon_{-}^{(1)}+\varepsilon_{+}^{(2)})}}\label{NN4:abss}.
 \eea
Moreover, from Eq. (\ref{NN4:2}) we have that a solution exists if
and only if the phases of the amplitudes satisfy the relation \be
\phi-\chi=m\pi,\quad m\in \mathbb{Z}\ee where $m$ must be
odd(even) for $(\varepsilon_{+}^{(1)}-
\varepsilon_{+}^{(2)})/(\varepsilon_{-}^{(1)}-\varepsilon_{-}^{(2)})$
negative(positive), respectively. Finally, it is
straightforward to show that the normalisation condition and the
non-negativity of the right-hand sides in Eqs. (\ref{NN4:abss})
imply that the intervals between $\left
[\varepsilon_{+}^{(1)},~\varepsilon_{+}^{(2)}\right ]$ and $\left
[\varepsilon_{-}^{(1)},~\varepsilon_{-}^{(2)}\right ]$ must overlap
but need not be one inside the other. For instance, one can readily
check that the spectra discussed in many papers
\cite{NPL04,CDEKL04,KS05,YB05} fulfil this condition and thus can
be obtained within the present formalism. We have also verified this
fact numerically for larger networks.

At this point, we have analysed completely the PST Hamiltonians
which lead to the permutation matrix (\ref{anti_perm}).
Indeed, using the expressions for the absolute values and choosing
the phases and the spectra according to the aforementioned constraints,
it is straightforward to derive possible combinations
of $\{E_1, E_2\}$ and $\{g_1, g_2\}$ which lead to PST transfer
by means of the equations
\[\bra{1}{\cal H}_{\bf l}^{(4)}\ket{1}=E_1,\quad
\bra{2}{\cal H}_{\bf l}^{(4)} \ket{2}=E_2,\]
and
\[\bra{1}{\cal H}_{\bf l}^{(4)}\ket{2}=g_1,\quad
\bra{2}{\cal H}_{\bf l}^{(4)}\ket{3}=g_2,\]
respectively.

The remaining question is whether there are other $4\times 4$
many-cycle permutation matrices generated by NN-type Hamiltonians.
To answer, we have to analyse the three remaining configurations of
cycles.
\begin{itemize}
\item Permutation $(4,1)(2)(3)$.
In this case we have two
cycles of lengths two, one, and one, respectively and two different
eigenvalues namely, $+1$ (degeneracy $\delta_{+1}=3$) and $-1$.
Following the theory of Sec. \ref{sec:manyCycles} and working as before
for the subspace $\cal{E}_+$, we can define new eigenvalues
$\varepsilon_+^{(i)}=2\pi l_+^{(i)}$ with the corresponding eigenvectors
in the computational basis given by
$| y_+^{(i)}\rangle=\mu_i (\ket{1}+\ket{4})/\sqrt{2}+\nu_i\ket{2}+
\xi_i\ket{3}$, for $i\in\{1,2,3\}$. In the subspace $\cal{E}_-$,
the eigenvalue and the eigenvector remain unchanged i.e.,
$\varepsilon_-=-1$ and
$|y_-\rangle= (\ket{1}-\ket{4})/\sqrt{2}$. Hence, the general
form of the Hamiltonian which leads to the permutation under
consideration after $\tau$ applications is given by
\[
{\cal \tilde{H}}_{\mathbf{l}}^{(4)}=\frac{1}{\tau}\left [
\sum_{i=1}^3\varepsilon_+^{(i)}\ket{y_+^{(i)}}\bra{y_+^{(i)}}
+\varepsilon_-\ket{y_-}\bra{y_-}\right ].
\]
Recall now that we are interested only in NN-type Hamiltonians i.e.,
Hamiltonians with $\bra{4}{\cal \tilde{H}}_{\bf l}^{(4)}\ket{2}=0$ and
$\bra{1}{\cal \tilde{H}}_{\bf l}^{(4)}\ket{3}=0$.
Rewriting ${\cal \tilde{H}}_{\mathbf{l}}^{(4)}$ in the computational
basis, we find that these two conditions also imply
$\bra{1}{\cal \tilde{H}}_{\bf l}^{(4)}\ket{2}=0$ and
$\bra{4}{\cal \tilde{H}}_{\bf l}^{(4)}\ket{3}=0$.
In other words, the first and the last sites are completely decoupled
from their respective neighbours and thus there are no NN-type
Hamiltonians which lead to PST in the context of the permutation
$(4,1)(2)(3)$. Strictly speaking, PST from the first to the last
site may occur only if these two sites are directly coupled.
This is, however, the case of a trivial two-site network for which PST
is always possible.

\item Permutation $(4,1,3)(2)$. In this case we have
two cycles of lengths three and one, respectively. The general form
of eigenvectors in the computational basis is $|y_+^{(1)}\rangle= \mu
(\ket{1}+\ket{3}+\ket{4})/\sqrt{3}+\nu\ket{2}$ and $|y_+^{(2)}\rangle=
-\nu^* (\ket{1}+\ket{3}+\ket{4})/\sqrt{3} +\mu^*\ket{2}$ for the
eigenvalue +1, $\ket{y_x}=(\ket{1}+x\ket{3}+x^2\ket{4})/\sqrt{3}$
for the eigenvalue $x$, and $\ket{y_{x^2}}=(\ket{1}+
x^2\ket{3}+x\ket{4})/\sqrt{3}$ for the eigenvalue $x^2$, where
$x=\exp(2{\rm i}\pi/3)$. Asking for a NN-type Hamiltonian, among
others, we obtain the constraints \bea
&&\varepsilon_{+1}^{(1)}|\mu|^2+\varepsilon_{+1}^{(2)}|\nu|^2+
\varepsilon_x x+\varepsilon_{x^2} x^2=0,\nonumber \\
&&\varepsilon_{+1}^{(1)}|\mu|^2+\varepsilon_{+1}^{(2)}|\nu|^2
+\varepsilon_x x^2+\varepsilon_{x^2} x=0,\nonumber \eea
which should
be satisfied simultaneously. This implies the condition
$(\varepsilon_x -\varepsilon_{x^2})(x-x^2)=0$ which cannot be
fulfiled because $\varepsilon_{x}$ and $\varepsilon_{x^2}$ are of
the form $\varepsilon_{x} = (2/3+2l_3)\pi$, and $\varepsilon_{x^2} =
(4/3+2l_4)\pi$, with $l_3,l_4\in \mathbb{Z}$.
\item  Permutation $(4,1,2)(3)$.
This case is similar to the permutation
$(4,1,3)(2)$.
\end{itemize}
Hence, for the four-site network, NN-type PST Hamiltonians involving
permutation transformation can be obtained only in the framework of
the antidiagonal permutation matrix (\ref{anti_perm}).

Concluding this simple example it is worth keeping in mind the
aforementioned observation about the overlap of the intervals
for odd and even eigenvalues. This turns out to be a particularly
useful result as it enables us to construct more PST
Hamiltonians by changing appropriately the spectrum of an already
known PST Hamiltonian. Most importantly, we have verified numerically
the functionality of this idea for interactions beyond nearest
neighbours and for networks involving more than four sites
(e.g., see Table \ref{tableCD} in the next example).
We turn now to discuss PST Hamiltonians beyond NN interaction.

\subsection{Beyond nearest-neighbour interaction}

In the previous example we explicitly discussed the
construction of NN-type PST Hamiltonians. In \cite{Kay06} the problem
of PST Hamiltonians beyond NN interaction was addressed for the
first time. Motivated by this work we have studied the problem of PST
Hamiltonians with a prescribed drop off of the coupling constants
also in our theoretical framework. In this case, naturally, we do have
only a very limited control over the form of the Hamiltonian.
However, before we proceed to present related results, we point out
that the class of Hamiltonians derived in Sec. \ref{Sec2.1} is
already of the beyond nearest neighbour type.

We have applied our method to Hamiltonians involving two different
types of interactions and for networks with a moderate number of
sites namely, four and six. For the latter case we present some of
the obtained numerical results. Consider the Hamiltonian of the form
\begin{widetext}
\bea
{\cal H}^{(6)}=\left(\begin{array}{cccccc} E_1 & \frac{1}{r_1^\gamma} &
\frac{1}{(r_1+r_2)^\gamma} & \frac{1} {(r_1+r_2+r_3)^\gamma} &
\frac{1}{(r_1+2r_2+r_3)^\gamma} & \frac{1}
{(2r_1+2r_2+r_3)^\gamma} \\
\frac{1}{r_1^\gamma} & E_2 & \frac{1}{r_2^\gamma} &
\frac{1}{(r_2+r_3)^\gamma}
& \frac{1}{(2r_2+r_3)^\gamma} & \frac{1}{(r_1+2r_2+r_3)^\gamma} \\
\frac{1}{(r_1+r_2)^\gamma} & \frac{1}{r_2^\gamma} & E_3 &
\frac{1}{r_3^\gamma}
& \frac{1}{(r_2+r_3)^\gamma} & \frac{1}{(r_1+r_2+r_3)^\gamma} \\
\frac{1}{(r_1+r_2+r_3)^\gamma} & \frac{1}{(r_2+r_3)^\gamma} &
\frac{1}
{r_3^\gamma} & E_3 & \frac{1}{r_2^\gamma} & \frac{1}{(r_1+r_2)^\gamma} \\
\frac{1}{(r_1+2r_2+r_3)^\gamma} & \frac{1}{(2r_2+r_3)^\gamma} &
\frac{1}
{(r_2+r_3)^\gamma} & \frac{1}{r_2^\gamma} & E_2 & \frac{1}{r_1^\gamma} \\
\frac{1}{(2r_1+2r_2+r_3)^\gamma} & \frac{1}{(r_1+2r_2+r_3)^\gamma}
& \frac{1} {(r_1+r_2+r_3)^\gamma} & \frac{1}{(r_1+r_2)^\gamma} &
\frac{1}{r_1^\gamma} &
E_1
\end{array}\right).
\label{DDI-ham} \eea
\end{widetext}
which refers to a network where interactions extend beyond nearest
neighbours and drop off with distance as $1/r^\gamma$. Our purpose
is to find combinations of energies $\{E_1,E_2,E_3\}$ and distances
$\{r_1,r_2,r_3\}$, such that ${\cal H}^{(6)}$ leads to an
antidiagonal $6\times 6$ permutation matrix after precisely $\tau$
applications.

As in the previous example, the Hamiltonian design proceeds in two
steps. First we have to find a general parameterisation of the
possible PST Hamiltonians along the lines of Sec.
\ref{sec:manyCycles}. Subsequently, asking for Hamiltonians of the
form (\ref{DDI-ham}), we obtain a set of equations for the free
parameters entering our design. This system of equations can be
solved by means of standard numerical techniques yielding a number
of acceptable solutions (i.e., sets of energies $\{E_1,E_2,E_3\}$
and distances $\{r_1,r_2,r_3\}$). A few of them, for the coulomb
$(\gamma=1)$ and the dipole-dipole $(\gamma=3)$ interaction are
given in Table \ref{tableCD}. In discussing this table, we would
like to note that the solution ($a$) for the dipole-dipole
interaction coincides (apart from a shift in the energies) with the
one given by Kay \cite{Kay06}, while all the other solutions are new
and have not been discussed in the literature before. Moreover, as
is evident from the table footnotes, for both solutions ($a$) and
($b$) we have an overlap of the spectrum for even and odd
eigenvalues. More solutions can be obtained in a straightforward
manner by modifying appropriately these spectra, while they overlap.

\begin{table}\caption{ Perfect state transfer. Parameters
entering the Hamiltonian (\ref{DDI-ham}), for $\gamma=1$ (Coulomb
interaction) and $\gamma=3$ (dipole-dipole interaction). The
depicted values for the energies and the distances should be
multiplied by $\pi/\tau$ and $(\tau/\pi)^{1/\gamma}$,
respectively.} \label{tableCD}
\begin{ruledtabular}
\begin{tabular}{cccccc}
&\multicolumn{2}{c}{Coulomb interaction}&&\multicolumn{2}{c}{Dipole
interaction}\\Parameter& Solution\footnotemark[1] & Solution\footnotemark[2] &
&Solution\footnotemark[1] & Solution\footnotemark[2]\\\hline
$E_1$& 1.07571   & 1.10848  & &   -0.00885  & 0.36328 \\
$E_2$& -0.70069  & -0.73164   & &  -0.61799  & -1.89782 \\
$E_3$& -1.87501  &  -3.87684  & &  -0.87315 & -1.96546 \\
$r_1$& 1.77185   &   1.93029   & &  0.96704  & 0.97927 \\
$r_2$& 1.12358   &   0.71338  & & 0.90156   & 0.74773  \\
$r_3$& 0.96133   & 1.06286   & &  0.88587   & 0.89090  \\
\end{tabular}
\end{ruledtabular}
\footnotetext[1]{Eigenenergies:
$\{\varepsilon_1, \varepsilon_2,\varepsilon_3,\varepsilon_4,\varepsilon_5,
\varepsilon_6\}=\{2, 0, -2, -1, -3 ,1\}$.}
\footnotetext[2]{Eigenenergies: $\{\varepsilon_1,\varepsilon_2,\varepsilon_3,
\varepsilon_4, \varepsilon_5,\varepsilon_6\}=\{2, 0, -4, -1, -5 ,1\}$.}\end{table}

Throughout this section, for the sake of simplicity and
illustration, we have focused on small networks and particular
choices of Hamiltonians. In closing we would like to emphasise that
the presented solutions are by no means exhaustive. Our theoretical
approach to the problem of PST is able to provide infinitely many
solutions and is not {\em a priori} restricted to particular
topologies and/or coupling configurations. In the following section
we turn to discuss the evolution of the excitation under the
influence of various types of PST Hamiltonians.

\section{Wave packet dynamics}
\label{Sec4}
The Hamiltonians given in the previous sections are quite
abstract. To illustrate in more detail the dynamics they can
induce we present a few examples of how the dynamics of a single
excitation will look like. In particular we will comment on two
parameters namely the occupation probability distribution and the
corresponding entanglement. As we mentioned before, there
are infinitely many PST Hamiltonians one may propose,
and each one of them may induce a new type of dynamics.
Hence, the patterns we have chosen to discuss in this section
at any rate cannot be considered as representatives
of the whole range of patterns generated by all the
possible PST Hamiltonians.

\subsection{Occupation probability distribution}\label{Sec4a}

In Sec. \ref{Sec2} we have explicitly constructed the whole class
of Hamiltonians $\Hr$ which satisfy condition (\ref{eq:1eq}). In
particular, we have seen that there are infinitely many
Hamiltonians ${\cal H}_\mathbf{l}$ which in general may lead to
different evolution operators and thus to different dynamics. A
more detailed classification is not an easy task. Nevertheless, in
this section we discuss typical transfer dynamics associated with
particular choices of the spectrum, or equivalently the integer
vector $\mathbf{l}$ entering the members of $\Hr$.

An excitation initially localized at site $i$ evolves under $m$
applications of the Hamiltonian (\ref{deg_Ham}) according to \bea
{\cal U}_\mathbf{l}(m)\ket{i}&\equiv& e^{{\rm i}{\cal
H}_\mathbf{l}m}\ket{i}\nonumber\\
&=&\sum_{\lambda_j\in\sigma}\sum_{k=0}^{\delta_{\lambda_j}}
\exp\Big ({\rm i}{\varepsilon_{\lambda_j}^{(k)}m}\Big )
\big|y_{\lambda_j}^{(k)}\big\rangle\big\langle
y_{\lambda_j}^{(k)}\big|i\big\rangle.
\nonumber\\
\eea The corresponding
probability to find the excitation at site $f$ is given by
\bea
P_f(m)&\equiv&\bra{f}{\cal U}_\mathbf{l}(m)\ket{i}
\nonumber\\
&=&\Bigg \vert
\sum_{\lambda_j\in\sigma}\sum_{k=0}^{\delta_{\lambda_j}}
\exp\Big ({\rm i}{\varepsilon_{\lambda_j}^{(k)}m}\Big ) \big\langle
f\big|y_{\lambda_j}^{(k)}\big\rangle \big\langle
y_{\lambda_j}^{(k)}\big|i\big\rangle\Bigg \vert^2.
\nonumber\\
\label{eq:opev} \eea Hence, during the transfer the probability of
finding the excitation may spread among all the sites of the
network which become essentially entangled. However, no matter how
complicated the transfer dynamics may be, we have the desired
perfect transfer of the excitation after exactly $\tau$
applications since ${\cal U}_{\mathbf{l}}(\tau)={\cal P}$.

Let us have a more detailed look at the single excitation (e.g.,
photon) wave packet dynamics, initially localised at the first
site i.e., $\ket{i}=\ket{1}$ in Eq. (\ref{eq:opev}). In Fig.
\ref{fig:occupy} we plot the evolution of the occupation
probability distribution as a function of the number of
applications $m$, for a medium size network of $11$ sites (modes).
The four plots differ by the choice of the spectrum. More
precisely, the first plot has been obtained for the spectrum with
$l_{\lambda_j} = 11-j$, the second plot is for the spectrum with
$l_{\lambda_{2j-1}} = 19 - 3 j$ and $l_{\lambda_{2j}} =0$ (for
$1\le j\le 6$), and the third plot corresponds to the symmetric
spectrum with $l_{\lambda_j} = |j - 6|$. The last plot corresponds
to an alternating sequence of eigenenergies with $l_{\lambda_{4
j-3}} = l_{\lambda_{4 j-2}}=0$, and $l_{\lambda_{4 j - 1}} =
l_{\lambda_{4 j}} = 5$, for $1\le j\le 3$.

The descendent sequence of eigenvalues used in Fig.
\ref{fig:occupy}(a) results in a gradual transfer of the
excitation from the initial position to the target position. At
each iteration instant the wave packet is slightly shifted and
only few of the adjacent modes is noticeable excited. On the
contrary, for all the other plots the wave packet is not only
broadened and propagating, but splits into two or more components.
As a result there is a nonzero probability to detect the
excitation at two or more distant sites at specific instances of
the evolution. In particular, when the decreasing sequence of
eigenvalues is interrupted by zero values, the probability
distribution pattern is formed by several ``paths'' which transfer
the initial excitation to the final target site [see Fig.
\ref{fig:occupy}(b)]. When the eigenvalue sequence exhibits a
symmetric dip, the probability distribution is formed by two
intersecting lines along which the probability propagates [see
Fig. \ref{fig:occupy}(c)]. The final plot [Fig.
\ref{fig:occupy}(d)], applicable for the tooth like sequence of
eigenvalues, shows an effective transfer of probability along the
outer parts of the network. The central modes \footnote{When
referring to ``central modes'' we do not refer to the topology of
the network but rather to the ordering of the computational basis
states.} are never significantly populated during the wave packet
propagation.

\subsection{Entanglement} \label{Sec4b}

\begin{figure*}
\includegraphics[width=12.cm]{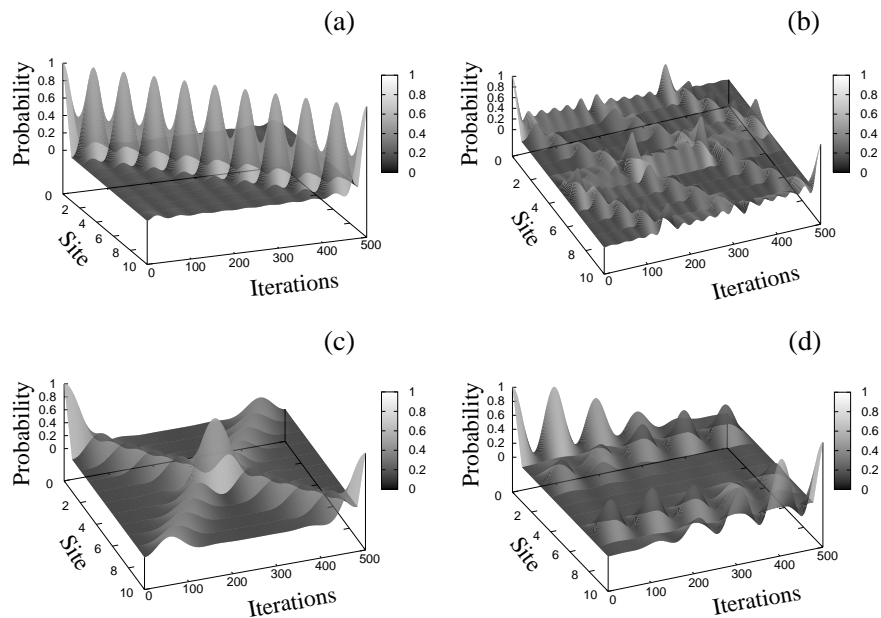}
\caption{Evolution of the occupation probability distribution
governed by PST Hamiltonians of various spectra.
Each plot corresponds to a network of eleven sites and $500$
successive applications of the corresponding PST Hamiltonian.
While the monotonously decreasing spectrum  induces a rather simple
diagonal pattern (a), the
other types of spectra induce more complicated patterns. Because
all the Hamiltonians involve in general non-vanishing couplings
 between the various sites, the wave packet can spread rapidly to
distant sites as can be seen from plots (b)-(d).}
\label{fig:occupy}
\end{figure*}

The behaviour of the single excitation (e.g., photon) in a network
implementing PST can be discussed also from the point of view of
entanglement. We can ask how the propagation of the photon in
different types of networks affects the mutual (bipartite)
entanglement. The bipartite entanglement can be quantified using
the concept of concurrence \cite{conc}. For the single photon
excitation the bipartite entanglement between two selected modes,
say $i$ and $j$, is given by \cite{single} \[C_{ij} (m) = 2
\sqrt{P_i(m) P_j(m)} .\] This simple result shows that shaping the
probability distribution strongly influences the entanglement
properties. The more the photon spreads across the modes during
the evolution the more entangled the modes become. When analysing
the bipartite entanglement between the initial mode and the
remaining modes we find that the entanglement plots exhibit a very
similar structure to the one of the probability distribution shown
in Figs. \ref{fig:occupy}(a-d). The main difference is that, due
to the probability multiplication in the definition of the
concurrence, the peaks become broader and less pronounced. Due to
the similarity we do not present any plots regarding the bipartite
entanglement. Instead, we proceed to discuss the single-excitation
dynamics from the point of view of the overall bipartite
entanglement $T (m)$, where additional conclusions can be drawn.

By definition, $T (m)$ is the sum of all $C^2_{ij}$ and is given
by\cite{single}
$$
T (m) = 2 \left[ 1 - \sum_{i} P_i (m)^2 \right] .
$$
Therefore, we see that the overall bipartite entanglement after
$m$ applications of the Hamiltonian is specified by the squared
probabilities (essentially the purity -- linearised entropy
related to the probability distribution). The more localized the
distribution is, the less overall bipartite entanglement the
propagating wave pattern contains. In particular for the patterns
presented in Figs. \ref{fig:occupy}(a-d) we come to the following
conclusions concerning the overall bipartite entanglement.

\begin{figure*}
\includegraphics[width=12.cm]{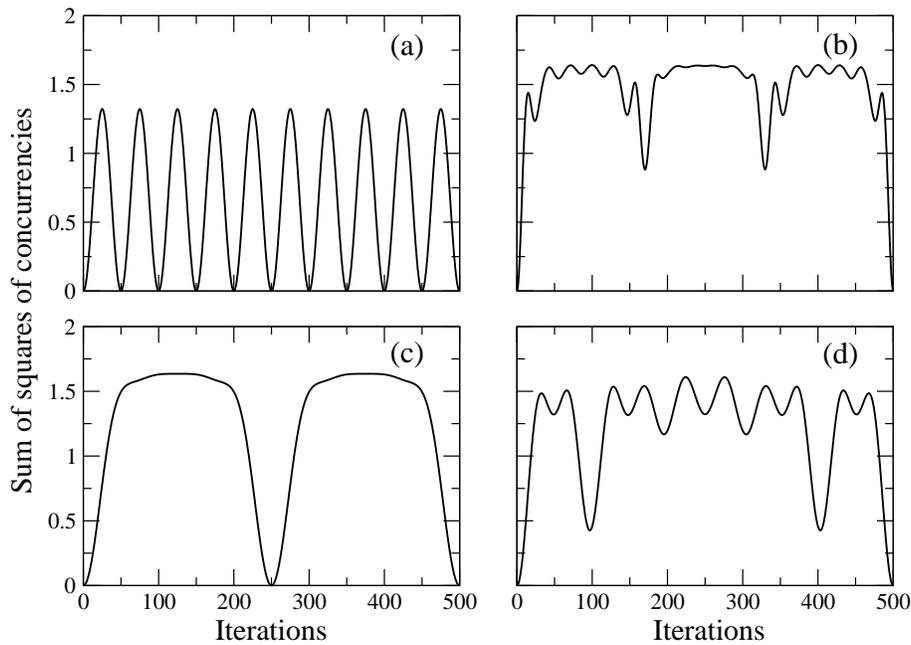}
\caption{
Evolution of the overall
concurrence corresponding to Figs. \ref{fig:occupy}(a)-(c),
as a function of iterations.
} \label{fig:ent}
\end{figure*}

The choice of the spectra influences the maximum value of $T$
which can be attained during the propagation as well as its
modulation. As is depicted in Fig. \ref{fig:ent}, the overall
bipartite entanglement for the monotonously decreasing [Fig.
\ref{fig:ent}(a)] and the symmetric-dip spectrum [Fig.
\ref{fig:ent}(c)] exhibits simple forms. In both cases the
excitation occupies at least one of the intermediate sites with
probability one [see Figs. \ref{fig:occupy}(a), (c) respectively]
and this fact is reflected in $T$ which drops to zero.

The overall bipartite entanglement for the two other spectra
acquires a more complicated structure and the maximum value of $T$
in Fig. \ref{fig:ent}(b) comes rather close to the maximum
achievable value of $1.81$ \cite{single}. This just indicates that
the wave packet is quite uniformly distributed among the sites
[see Fig. \ref{fig:occupy}(b),(d)]. The question whether the
prescribed maximum can be indeed reached by some Hamiltonian for
any of the classes we have discussed throughout this work is not
difficult to answer. It is sufficient to take as an example the
Hamiltonian inducing the Fourier transform. This transform is
cyclic and its square leads to a permutation matrix, hence is
among the Hamiltonians described by our method. At prescribed
instances it distributes the excitation uniformly among the sites
and hence will saturate $T$ for any number of sites.

\section{Conclusions}
\label{Sec5}

We analysed the problem of PST through passive networks(channels)
and the Hamiltonians which induce the corresponding transform. The
transform is represented by a permutation  matrix which has
particularly simple algebraic properties. Hence, in the case of
one-cycle permutations we were able to obtain closed expressions
for the whole class of NN-type Hamiltonians, while we proved that there
exists no PST Hamiltonian in the framework of NN interactions. In
the case of many-cycle permutations,  the situation is more
involved and for the time being the derivation of analytic expressions
seems to by possible only in the case of relatively small networks.
However, at any rate the problem can be still treated  numerically.

In contrast to previous work in the field, our theory is not
limited by any {\em a priory} restrictions to the topology of the
system and the configuration of couplings between different sites
of the network. Hence, our approach provides new ways for quantum
wire engineering for systems of arbitrary topology and
interactions. In this spirit, we were able on the one hand to
recover all the known types of PST Hamiltonians restricted to
centrosymmetric networks and/or NN interaction, and on the other
hand to derive new Hamiltonians. In principle there are infinitely
many PST Hamiltonians which can be described within our
theoretical framework. However, we would like to emphasise that
the different types of Hamiltonians need not be always
implementable by a specific experimental setup. Actually, which
types of Hamiltonians are indeed implementable has to be decided
on the basis of the types of the particular physical interactions
used in the system. Finally, apart from studying the form of PST
Hamiltonians we also discussed the dynamics induced by them in
terms of occupation probability distribution and overall
bipartite entanglement.

Additional interesting questions linked to the present problem
such as, stability of the Hamiltonians with respect to coupling
constant perturbation, detailed analysis of the structure of the
permutation cycles and the form of corresponding Hamiltonians,
generalisations to higher dimensions etc, are left for further
investigation.

\section{Acknowledgement}
Financial support by GA\v CR 202/04/2101, by the DAAD (GA\v CR
06-01), EU QUELE and by the projects LC 06002 and MSM 6840770039 of the Czech Ministry
of Education is gratefully acknowledged. GMN also acknowledges
support by ``Pythagoras II'' of the EPEAEK research programme.

\end{document}